\documentclass[12pt]{article}
\usepackage{epsf,latexsym,indentfirst}
\input{epsf}

\def\be{\begin{equation}}
\def\ee{\end{equation}}
\def\bea{\begin{eqnarray}}
\def\eea{\end{eqnarray}}

\author{
A. Patk\'os$^1$\footnote{patkos@ludens.elte.hu}\,,
P. Petreczky$^{1,2}$\footnote{petr@hercules.elte.hu\vskip0truecm
\hspace{0.28cm}petreczk@Physik.Uni-Bielefeld.DE}\,\,,
Zs. Sz\'ep$^1$\footnote{szepzs@hercules.elte.hu} \\
\makebox[0.5cm]{}$^1$Dept. of Atomic Physics, E\"otv\"os University,\\
\qquad H-1088 Puskin 5-7, Budapest Hungary \\
$^2$Fakult\"at f\"ur Physik, Universit\"at Bielefeld,\\
\qquad\qquad  P.O. Box 100131, D-33501 Bielefeld, Germany
}
\title{
\small
\begin{flushright}
ITP Budapest Rep. 537 \\
BI-TP 97/45 \\
November, 1997
\end{flushright}
\vspace {1.5cm}
\Large  \bf Coupled Gap Equations for the \\
Screening Masses in Hot SU(N) Gauge Theory}
\normalsize
\date{\vskip1.5cm}

\begin{document}
\maketitle
\thispagestyle{empty}
\begin{abstract}
\noindent
Coupled 1-loop gap equations are studied numerically for non-Abelian
electric and magnetic screening in various versions of the
three-dimensional effective gauge models. Corrections due to higher
dimensional and non-local operators are assessed quantitatively.
Comparison with numerical Monte-Carlo investigations suggests that
quantitative understanding beyond the qualitative features can be
achieved only by going beyond the present treatment. 
\end{abstract}
\newpage
\pagenumbering{arabic}

\section{Introduction}
An outstanding consequence of heating non-Abelian gauge fields is the screening
of static chromo-electric and -magnetic fields. Electric (Debye) screening is 
generated by non-zero Matsubara modes both for Abelian and non-Abelian 
gauge fields and its value at leading order has been known since a long time. 
The dynamics of the non-Abelian zero modes is quite complex \cite{lin80}
and it leads, at least in the Schwinger-Dyson approach to the static magnetic 
gluon two-point function, to the generation of a magnetic screening 
mass \cite{phil94}. 
Non-zero magnetic mass was obtained from all variants of the magnetic 
gap equation in the effective three-dimensional gauged Higgs and also in 
pure gauge theories \cite{phil94,buch93,espin93,buch94,alex95}. 

The spectra of screening masses in the SU(2) Higgs model were also
studied in lattice Monte-Carlo simulations by measuring various correlation
functions of gauge invariant operators 
\cite{kaj96, phil96,gurt96} as well as the gauge boson propagator
in fixed (Landau) gauge \cite{karsch96}. Electric and magnetic screening
are simultanously accessible to four-dimensional Monte-Carlo simulations
\cite{karsch97}. The magnetic mass obtained from the gauge boson propagator in 
Landau-gauge is rather close to that obtained from the gap equation, while
the gauge invariant gluonic correlation functions yield several times larger
mass. A consistent interpretation of the situation emerging
in the weakly coupled Higgs+Gauge model was suggested in
\cite{buch96} (see also \cite{dosch96}) in the framework of the so-called 
constituent model, where the mass measured in fixed gauge correlators 
corresponds to the constituent magnetic gluon mass of the confined 
three-dimensional theory. The idea of the constituent gluon offers
a natural effective degree of freedom, which dominates the static part
of the free energy of the gluon plasma \cite{pat2}. It should be noticed
in this context that the relation between gap equation and the
resummation of the free energy has been  recently discussed in
\cite{petr1,rein,dru} for the scalar field theory.

For pure hot $SU(N)$ theory the situation is more complicated since for
the physically interesting range of the temperature the
value of the gauge coupling
$g$ is close to unity and it is not clear whether the standard
argument for dimensional reduction applies. 

The aim of the present paper is to investigate how far the screening
masses can be described by means of the technique of gap equations.
Since the hiearchy of scales $2 \pi T>>g T>>g^2 T$ fails to hold
in pure $SU(N)$ theory for the temperature range of interest we
should investigate a coupled set of gap equations for all the screened 
modes and determine the corresponding screening lengths simultanously. 

The most straightforward way to do this would be to
derive gap equations in the full four-dimensional theory. However, it is
not known how to generalize the by-now well-established
three-dimensional gauge invariant resummation \cite{buch94,alex95} to 
four dimensions. There are gauge transformations which might mix
static and non-static modes, therefore the resummation of the
static modes only,
which was suggested in \cite{arn93} violates gauge invariance.
Since screening masses are static quantities it is natural to
calculate them in the framework of an effective three-dimensional
theory which is, however, valid only up to scales $k\simeq g T$. It was
shown that the accuracy of the description of such theories is improved if in 
the action of the effective theory beside superrenormalizable
operators one also takes into account higher dimensional and
non-local operators \cite{jak96}. When deriving the gap equation in the
framework of three-dimensional adjoint Higgs model it is important  to
address the question whether the symmetric or the broken phase is the
physical one. In Refs. \cite{kaj96,kark} it has been argued that symmetric
phase is the physical one, the study of the Debye screening also suggests
that the presence of the $A_0$ condensate is physically unfavourable,
therefore in what follows we will assume no $A_0$ backround.

In this paper we shall assume only the separation of the static and 
non-static scales in the pure $SU(N)$ gauge theory, that is we assume
that $g(T)<<2\pi$. In this way we will be interested in the derivation
and the solution of the coupled gap equations of the electric and magnetic
static fields. To our best knowledge our paper represents the first attempt
for this simple generalisation of the gap-equation approach, where till
now electric and magnetic screening masses have been discussed separately. 
The closest to the spirit of our investigation is the analysis of the 
sensitivity of higher order corrections of the electric screening mass to the
existence and the value of a magnetic screening scale by Rebhan \cite{reb94}.

The paper proceeds as follows: in section 2 we shall calculate screening
masses from the local superrenormalizable effective theory (i.e the adjoint 
Higgs model) of the hot $SU(N)$ gauge theory using a gauge-dependent resummation
scheme (namely, $R_\xi$-gauge). The temperature
and gauge-parameter dependence of the results are discussed in detail.
Gauge invariant resummation schemes are used for the derivation of 
similar coupled gap-equations in section
3. In section 4 we shall analyze  the effect of higher dimensional and
non-local operators in the gap equation. In section 5 detailed comparison
with the results of earlier and the most recent numerical investigations will 
be presented together with our conclusions.

\section{Gauge non-invariant resummation scheme}
This scheme for the evaluation of the magnetic mass was first
suggested in \cite{phil94} and for the Debye mass in
\cite{reb94}. It was noticed in Ref. \cite{phil94} that magnetic mass 
obtained in this scheme is gauge dependent and therefore cannot be regarded as 
physically meaningfull. However, even in gauge invariant resummation schemes the 
value of the magnetic mass cannot be defined unambigously, because it depends 
on the specific resummation scheme \cite{jac97}. Therefore it seems 
interesting to calculate the magnetic mass in a gauge non-invariant
scheme just to compare the amount of gauge dependence with the ambiguity of 
gauge invariant approaches.

The three-dimensional effective Lagrangian relevant for 1-loop
calculations can be written with $R_\xi$ gauge fixing as
\bea
&
L={1\over 4}F^a_{ij} F^a_{ij}+{1\over 2}\left({(D_i A_0)}^2+m_D^2 A_0^a A_0^a
\right)+
\nonumber\\
&
{1\over 2} m_T^2 A^a_i A^a_i+(\partial_i \bar c^a D_i c^a+
m_G^2 \bar c^a c^a)+
{1\over 2 \xi} {(\partial_i A_i)}^2+L_{ct},\nonumber\\
&
L_{ct}={1\over 2}(m_{D0}^2-m_D^2)A_0^a A_0^a-{1\over 2} m_T^2 A_i^a A_i^a-
m_G^2 \bar c^a c^a\\\nonumber
\textrm{with}&\makebox[14.9cm]{}&\\
\nonumber
&D_iA_{0}^{a}=\partial_i A_0^a-g_3 f^{abc}A_0^b A_i^c,\\\nonumber
&F_{ij}^a=\partial_i A_j^a-\partial_j A_i^a+g_3f^{abc} A_i^bA_j^c.
\eea

Here we have added and substracted a  mass term for $A_i$, $A_0$ and the
ghost fields with their exact screening masses. $m_{D0}^2$ is the tree-level 
(from the point of view of the effective theory) Debye mass for $A_0$, which
was generated during the procedure of the dimensional reduction.
The other parameter of the effective theory is the three-dimensional gauge
coupling $g_3$ which appears in the definitions of the covariant
derivatives and the field strength tensor. At 1-loop level one has 
$m_{D0}^2=g^2 N T^2/3$ and $g_3^2=g^2 T$, where $g=g(T)$ is the gauge
coupling of the original theory. The propagators can be read from the
quadratic part of the Lagrangian and can be found in the Appendix, where one
also finds some details of the evaluation of the relevant Feynman diagrams 
contributing to the different 2-point functions. It should be noticed
when performing the resummation of the pure gauge sector with the unique
mass term of mass $m_T$ the longitudinal and transverse gluons
acquire different masses, which are, however related by
$m_L=\sqrt{\xi} m_T$ ($m_T$ is the transverse and $m_L$ is the
longitudinal mass). It is also possible to perform the resummation by
introducing independent masses for the longitudinal and transverse gluons,
but then the corresponding gap equations will have only complex
solutions.
The gauge boson self-energy can be decomposed as
\be 
\Pi_{ij}(k)=(\delta_{ij}-{k_{i} k_{j}\over k^2})
\Pi_T(k,m_T,m_D,m_G)+{k_{i} k_{j}\over k^2} \Pi_L(k,m_T,m_D,m_G).
\ee
In the Appendix we give the expression of $\Pi_{ij}$ in terms of a few
fun\-da\-men\-tal three-dimensio\-nal loop-integrals. These integrals are 
easily evaluated and an explicit but very cumbersome functional form can be 
written for the longitudinal and transversal projections of the polarisation 
matrix.

The self energy for $A_0$ was first calculated in \cite{reb94}:
\bea
&
\Pi_{00}(k,m_D,m_T)=m^{2}_{D0}+{g^2 N\over 4 \pi} \biggl[-m_D-m_T+
{2 (m_D^2-k^2-m_T^2/2)\over k} \arctan{k\over m_D+m_T}\nonumber\\
&
+(k^2+m_D^2) \biggl({k^2+m_D^2\over m_T^2 k} (\arctan{k\over
m_D+\sqrt{\xi} m_T}-\arctan{k\over m_D+m_T})+(\sqrt{\xi}-1){1\over
m_T}\biggr)\biggr].
\label{reb}
\eea
\begin{figure}
\epsfbox[-90 0 100 200]{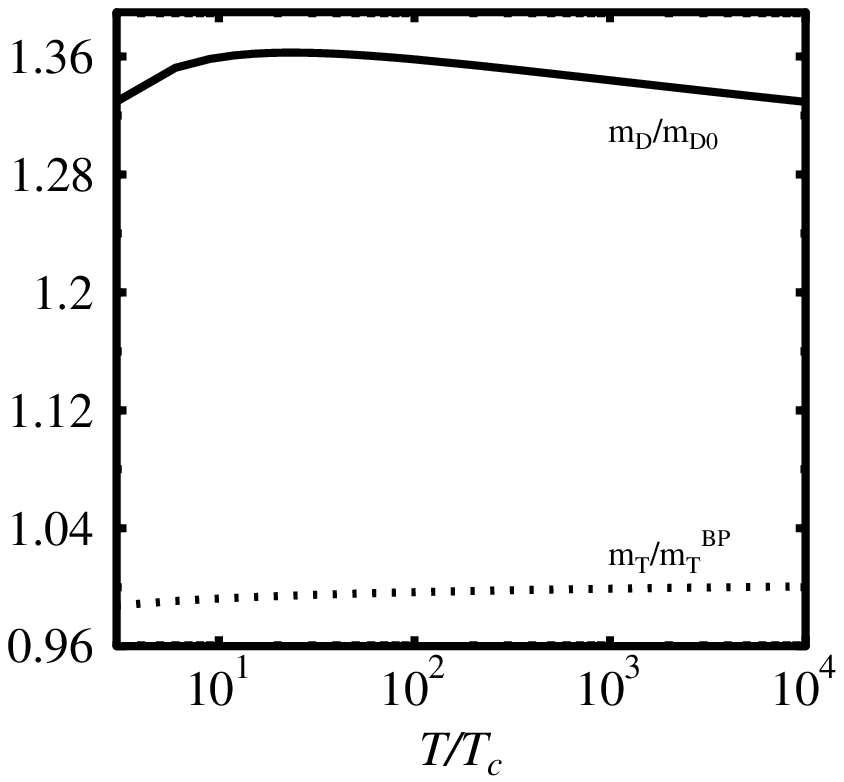}
\caption{The temperature dependence of $m_D/m_{D0}$ (solid line) and
$m_T/m_T^{BP}$ for Feynman ($\xi=1$) gauge, $m_{D0}$ is  the leading 
order result for the Debye mass and $m_T^{BP}$ is the value of the
magnetic mass obtained by Buchm\"uller and Philipsen for pure $SU(2)$
gauge theory.}
\vskip 2truecm
\epsfbox[-90 0 100 200]{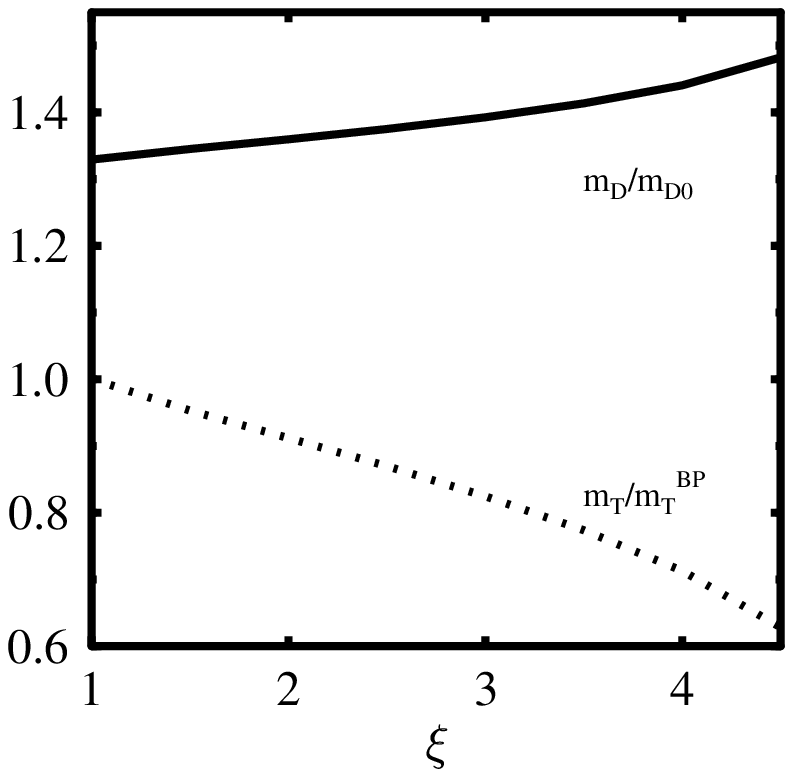}
\caption{The dependence of $m_D/m_{D0}$ (solid line)
and $m_T/m_T^{BP}$ on  gauge parameter $\xi$ at $T=10^4 T_c$, $m_{D0}$ 
is the leading order result for the Debye mass and $m_T^{BP}$ is the value 
of the magnetic mass obtained by Buchm\"uller and Philipsen for pure $SU(2)$
gauge theory.} 
\end{figure}
The on-shell gap equations now can be written as
\begin{eqnarray}
\nonumber
m_T^2 &=&\Pi_T(k=im_T,m_T,m_D,m_G), \\
m_D^2 &=&\Pi_{00}(k=im_D,m_D,m_T), \\\nonumber
m_G^2 &=&\Pi_G(k=i m_G,m_T,m_G). 
\end{eqnarray}
On the mass-shell $\Pi_{00}(k=i m_D,m_D,m_T)$ is gauge parameter
independent, but  $\Pi_T(k=i m_T,m_T,m_D,m_G)$ and $\Pi_G(k=i
m_G,m_T,m_G)$ do depend on the gauge fixing parameter, therefore the
masses obtained from this coupled set of gap equations are gauge dependent. 

In the following numerical investigations we shall consider the case
of the $SU(2)$ gauge group. The 4-dimensional coupling constant is taken
at scale $\bar \mu=2 \pi T$, where $\bar \mu$ is the $\overline{MS}$ scale and
1-loop relation for the gauge parameter of the effective theory is used. To set 
the temperature scale we use the relation $T_c/{\Lambda_{\overline{MS}}}=1.06$
obtained from numerical simulation of the finite temperature $SU(2)$ gauge 
theory \cite{karsch97}.

The temperature dependence of $m_D$ in a specific gauge  is plotted in 
Fig.1. As one can see from the plot $m_D$ receives $30 \%$ positive
correction compared to the leading order result, while the
magnetic mass stays very
close to the value calculated by Buchm\"uller and Philipsen \cite{buch94}, given
below in Eq.(\ref{egy6}). Since the masses are gauge dependent in this
approach, it is important to investigate the dependence of the screening masses 
on the gauge parameter $\xi$. It turns out that one gets real values for the 
masses from the gap equations only if $\xi\in[1,5)$. The dependence of the 
screening masses on $\xi$ in this range is shown in Fig.2. One can see, that
the $\xi$-dependence of $m_T$ in this interval is 40 \%, while for $m_D$ it
remains in the 10 \% range.

\section{Gauge Invariant Approach}

Gauge invariant approaches  for the magnetic mass
generation in three-dimensional pure $SU(N)$ gauge theory  were
suggested by Buchm\"uller and Philipsen (BP) \cite{buch94} and by
Alexanian and Nair (AN) \cite{alex95}. 
The approach of AN uses the hard thermal loop inspired effective action
for the resummation of the magnetic sector. The approach of BP using a
gauged $\sigma$-model, goes over to the $SU(N)$ gauge theory in the limit of
infinitely heavy scalar field. Till now only these two gauge invariant
schemes are known to provide real values for the magnetic mass \cite{jac97}.
 
In these approaches the gauge boson self-energy is automatically
transverse and there is no need to project the transverse part from the 
polarisation tensor. The corresponding expression for the on-shell
self-energy reads
\be
\Pi_T(k=i m_T,m_T)=C m_T,
\ee
where 
\be
\label{egy6}
C=\cases{{g^2 N\over 8 \pi} [{21\over 4} \ln 3-1],AN ,\cr
{g^2 N\over 8 \pi} [{63\over 16} \ln 3-{3\over 4}],BP.}
\ee

Since we are interested in calculating the screening masses in the 
three-dimensional $SU(N)$ adjoint Higgs model, $\Pi_T(k,m_T)$ should be
supplemented by the corresponding contribution coming from $A_0$
fields. This contribution is calculated from diagrams d) and e) of the
Appendix to be
\be
\delta \Pi_{ij}^{A_0}(k,m_D)=
{g^2 N\over 4 \pi} \left(-{m_D\over 2}+{k^2+4 m_D^2\over 4 k}
\arctan{k\over 2 m_D}\right) \left(\delta_{ij}-{k_{i} k_{j}\over k^2}\right).
\ee
It is transverse and gauge parameter independent, it also does
not depend on the specific resummation scheme applied to the
magnetostatic sector. It should be also noticed that it starts to contribute to 
the gap equation  at ${\cal O}(g^5)$ level in the
weak coupling regime, thus preserving the magnetic mass scale to
be of order $g^2 T$. This is the reason why no "hierarchy" problem arises in
this case, at least for moderate g values.

\begin{figure}
\epsfbox[-90 0 100 200]{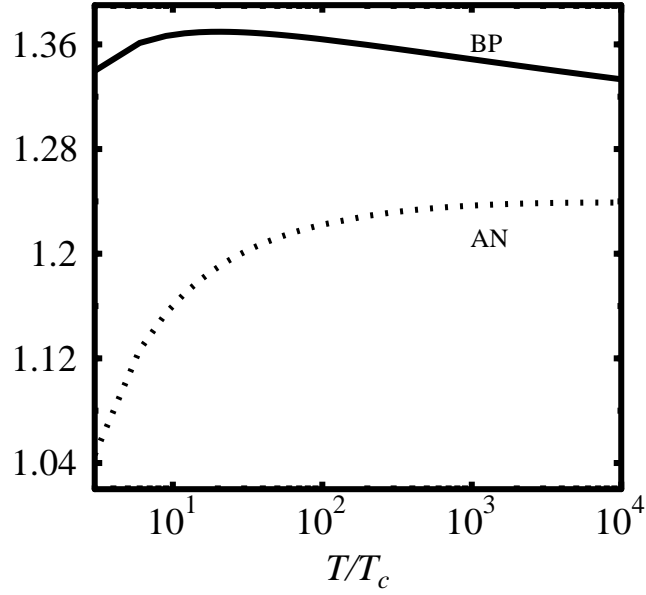}
\caption{The temperature dependence of the scaled Debye mass for
BP resummation scheme (solid) and for the AN resummation scheme
(dashed). The scaling factor is $m_{D0}$.}
\end{figure}
\begin{figure}
\vskip 2truecm
\epsfbox[-90 0 100 200]{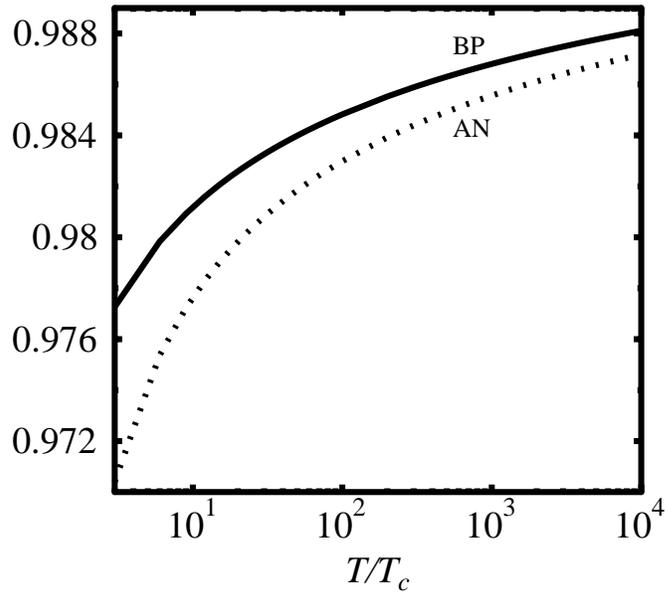}
\caption{The temperature dependence of the scaled magnetic  mass for
BP resummation scheme (solid) and for the AN resummation scheme
(dashed). The scaling factors are $m_{T}^{BP}$ and $m_{T}^{AN}$,
respectively.}
\end{figure}

The self energy of $A_0$ depends on the specific resummation scheme. 
For BP resummation it reads
\bea
&
\Pi_{00}(k,m_D,m_T)=m^{2}_{D0}+{g^2 N\over 4 \pi} \biggl[-m_D-m_T+
{2 (m_D^2-k^2-m_T^2/2)\over k} \arctan{k\over
m_T+m_D}-\nonumber\\
&
{(k^2+m_D^2)\over m_T^2} \biggl(-m_T+{1\over k} \arctan{k\over m_T+m_D}
\biggr)\biggr].
\eea
This expression is different from the expression of $\Pi_{00}$ calculated in 
the gauge non-invariant approach (see eq. (\ref{reb}) ) but its
analytic properties and on-shell value is the same as of (\ref{reb}).
For the resummation scheme of AN the self-energy expression of $A_0$
coincides with (\ref{reb}) if it is evaluated at $\xi=1$.
The coupled set of gap equations now can be written as
\bea
\nonumber
m_T^2&=&C m_T+\delta \Pi^{A_0}(k=i m_T,m_D), \\
m_D^2&=&\Pi_{00}(k=i m_D,m_D,m_T).
\eea
The temperature dependence of $m_D$ obtained from this coupled set of gap 
equations is shown in Fig.3 for both schemes, where we have again normalized the 
Debye mass by the leading order result, $m_{D0}$. The temperature dependence of
the magnetic  mass is shown in Fig.4, where we have normalized $m_T$ by the 
value of the magnetic mass obtained for pure three-dimensional $SU(2)$ theory, 
in the BP (AN) gauge invariant calculations \cite{buch94,alex95}. As 
one can see the contribution of $A_0$ to the magnetic mass is between 1 and 3\%. 
From Figures 3 and 4 it is also seen that the temperature dependence of the 
screening masses is very similar to the temperature dependence of the
respective leading order results.

\section{Contribution of non-Local Operators to the Gap\\ Equation}
In the previous section the gap equations were derived for an effective
local superrenormalizable theory. In this case the effect of non-static modes
in the 2-point function were represented by the thermal mass for $A_0$ and
by the field renormalization factors which relate 3d fields to the
corresponding 4d ones. It was shown in \cite{red,kajrum94} 
that when one performs the procedure of the dimensional
reduction in $R_{\xi}$ gauges  the parameters of the superrenormalizable
effective theory are gauge independent, only the expressions of 3d
fields in terms of 4d ones depend on the gauge parameter. However, this
does not hold for higher dimensional and non-local operators, which are
generally gauge dependent. At 1-loop level the only diagrams contributing
non-locally to the gap equation are those which have two non-static line
inside the loop, diagrams with one static and one non-static line inside
the loop are forbiden because of 4-momentum conservation. Therefore at
1-loop level the non-locality scale $(2\pi T)^{-1}$ is much smaller than the
relevant length scales.

In general the non-static contribution to the static 2-point function 
$\Pi_{\mu \nu}(k_0=0,k)$ can be written as
\bea
&
\Delta \Pi^{ns}_{\mu \nu}(k_0=0,k)=
\delta^0_{\mu} \delta_{\nu}^0 \Pi^{ns}_{00}(k)+\delta_{\mu}^i \delta_{\nu}^j (\delta_{ij}-
{k_i k_j\over k^2}) \Pi(k)
\nonumber, \\
&
\Pi_{00}^{ns}(k)=m_{D0}^2+a_1(\mu,\xi)k^2+T^2 \sum_{n=2}^{\infty}a_n(\xi)
 {({k^2\over 
2 \pi T})}^{2 n},\nonumber\\
&
\Pi^{ns}(k)=b_1(\mu,\xi) k^2+T^2 \sum_{n=2}^{\infty}b_n(\xi) {({k^2\over 
2 \pi T})}^{2 n},
\label{nonloc}
\eea
where the coefficents $a_n$ and $b_n$ can be calculated for arbitrary $n$.
The first two terms in the expressions of $\Pi_{00}$ and first term of
$\Pi (k)$ are already included into the 3d 
effective theory as part of the tree level mass and the
definition of the 3d fields in
term of 4d fields. The last two sums will contribute to the 3d effective
lagrangian as quadratic non-local operators. There are also higher dimensional
operators as well as non-local 3- and 4-point verticies in the effective
lagrangian, however, since we restrict our interest to 1-loop gap equations 
these are not important for us. Their contribution  would correspond to 2 or 
higher loop contribution in the full four-dimensional theory.
 
We have estimated by direct numerical evaluation of the infinite sums
in (\ref{nonloc}) the contribution of non-local operators to be less
than 1\% in the temperature range $T=\left(3-10^4\right)T_c$, thus neither 
their contribution nor their gauge dependence is essential.

\section{Conclusion}
In the present paper we have made an attempt to extract the electric and
magnetic screening masses from the coupled set of gap equations of the 
three-dimensional $SU(N)$ adjoint Higgs model considered as an effective theory 
of QCD. The screening masses have been studied using gauge non-invariant as
well as gauge invariant resummation schemes. In the gauge non-invariant
formalism we have observed rather strong gauge parameter dependence,
therefore the results extracted from it are not very informative.
It is still interesting to note that in Feynman gauge ($\xi=1$) the
results for the magnetic mass
are rather close to those obtained from the gauge invariant
resummation scheme of Buchm\"uller and Philipsen. In gauge invariant treatments 
we have compared two different resummation schemes, that of Buchm\"uller and 
Philipsen (BP) and one proposed by Alexanian and Nair (AN). Qualitativelly these 
two resummations lead to similar results, but in the BP scheme one has
smaller magnetic mass and larger Debye mass than in the AN scheme.

Let us summarize our view on the interaction of the electric $A_0$ and the
magnetic $A_i$ fields. In both schemes one can see that the dynamics of $A_0$ is
largely influenced by the magnetic sector, however, no similar feedback on the 
magnetic sector is seen, the magnetic masses calculated from the coupled gap 
equations provide screening masses which are 1\% smaller than evaluated in the 
pure gauge theory. This fact suggests that the role of the adjoint scalar field
is similar to that of the fundamental Higgs field, because the magnetic mass 
calculated in the symmetric phase of $SU(2)$ Higgs theory using lattice 
Monte-Carlo simulation with Landau gauge fixing is also
roughly the same as in pure gauge theory \cite{karsch96}. 

Finally we compare our results with recent Monte-Carlo data for the screening 
masses obtained in 4d finite temperature $SU(2)$ gauge theory \cite{karsch97}. 
The data on the magnetic mass found from this simulation in the temperature 
range $T=(10-10^4)T_c$ can be fitted well by the formula $m_T=0.456(6) g^2 T$. 
This value of the magnetic mass is considerably larger than what one obtains 
from the magnetic gap equation and 3d simulations, where the results are 
approximately $m_T=0.28 g^2 T$ for the BP scheme, $0.38g^2 T$ for AN scheme 
and $0.35 g^2 T$ for 3d simulation. Also our results are rather close
to these values, as we have demonstrated in Sections 2 and 3.
 
The data from Monte-Carlo simulation for the  Debye mass in the above mentioned 
temperature interval can be fitted using the following leading order-like  
anzatz $\sqrt{1.69(2)} g(T) T$ \cite{karsch97}, which means that the Debye mass 
is  roughly $1.6 m_{D0}$, where $m_{D0}$ is the leading order result.
The gap equations at the same time, as one can see from Fig.3 give  
$(1.2-1.3) m_{D0}$, depending on the resummation scheme. While there is no 
quantitative agreement between masses measured in Monte-Carlo simulation and 
those obtained from the gap equation, the temperature dependence of these masses 
in the temperature interval $T=(10-10^4)T_c$ seems to follow the temperature 
dependence of the leading order result.

At the moment we have no explanation for the reasons of the discrepancy between 
the results of the 4d simulation and the 3d gap equations and simulation. The
observations may imply that either higher loop contributions to the gap equation 
are important and therefore the dynamics of $A_0$ is not faithfully reflected by 
1-loop gap equations or the dimensional reduction is not valid.
In \cite{corn} the self-consistency of the gap equation has been examined in
the entire momentum space in gauge invariant way. This
investigation reveals some inconsistencies of the 1-loop gap equation, which
are conjectured to be removed by higher order calculation. It remains an open 
question how far these inconsistencies influence the pole mass.
We believe, 
however, that further studies in this direction, which include 2-loop gap 
equations \cite{eberline} and lattice study of the effective adjoint Higgs 
model \cite{petr} will clarify part of these problems.
\vskip1truecm
{\large \bf Acknowledgment}
\vskip0.5truecm
P.P. thanks W.Buchm\"uller, Z.Fodor, A.Jakov\'ac and F.Karsch for useful
discussion. P.P. was supported by Peregrination II. Found and partially
througth the ZiF project "Multi-Scale Phenomena and their Simulation".
A.P. thanks support from OTKA, grant No. T22929.
We thank K.Kajantie for reading the preprint version of the present paper
and for making some useful comments.

\section*{Appendix}

The propagators of different fields can be read off the quadratic part of
the Lagrangian. These exact propagators are listed below.
The gauge boson propagator:
\be
D_{ij}(k)=\left(\delta_{ij}-{k_i k_j\over k^2}\right){1\over k^2+m_T^2}+{k_i
k_j\over k^2}{\xi\over k^2+ m_L^2},
\ee
where $m_L=\sqrt{\xi} m_T$,
the propagator for the adjoint scalar field $A_0$:
\be
D^{A_0}(k)={1\over k^2+m_D^2},
\ee
and finally, the ghost propagator:
\be
\Sigma(k)={1\over k^2+m_G^2}.
\ee

The Feynman diagrams contributing to the 2-point functions of the relevant
fields are shown below. In these diagrams every line corresponds to a
resummed propagator. The wavy line corresponds to the gauge
bosons, the solid one to $A_0$ and the dashed one to the ghost.
The corresponding analytic expressions can be written in terms of the
following standard integrals:
\begin{eqnarray}
I(k,r,s)&=&\int\frac{d^d p}{(2\pi)^d}\frac{1}{(p+k)^{2r}p^{2s}}\nonumber\\
&=&\frac{k^{d-2(r+s)}}{(4\pi)^{d/2}}\frac{\Gamma\left(r+s-\frac{d}{2}\right)}
{\Gamma(r)\Gamma(s)}\,\frac{\Gamma\left(\frac{d}{2}-s\right)
\Gamma\left(\frac{d}{2}-r\right)}{\Gamma(d-s-r)},\\
I^{m}(k,r,s)&=&\int\frac{d^d p}{(2\pi)^d}\frac{p_{m}}{(p+k)^{2r}p^{2s}}
\nonumber\\
&=&-k^{m}\,\frac{k^{d-2(r+s)}}{(4\pi)^{d/2}}\frac{\Gamma\left(r+s-\frac{d}{2}
\right)}{\Gamma(r)\Gamma(s)}\,\frac{\Gamma\left(\frac{d}{2}+1-s\right)
\Gamma\left(\frac{d}{2}-r\right)}{\Gamma(d+1-s-r)},\\
I^{mn}(k,r,s)&=&\int\frac{d^d
p}{(2\pi)^d}\frac{p^{m}p^{n}}{(p+k)^{2r}p^{2s}}
\nonumber\\
&=&\frac{k^{d-2(r+s)}}{(4\pi)^{d/2}}\biggl[\frac{k^2}{2}\frac{\Gamma\left(r+s-1-
\frac{d}{2}\right)}{\Gamma(r)\Gamma(s)}\,\frac{\Gamma\left(\frac{d}{2}+1-s\right)
\Gamma\left(\frac{d}{2}+1-r\right)}{\Gamma(d+2-s-r)}\enspace\delta^{mn}
\nonumber\\
&&\qquad\qquad+\frac{\Gamma\left(r+s-\frac{d}{2}\right)}{\Gamma(r)\Gamma(s)}\,
\frac{\Gamma\left(\frac{d}{2}+2-s\right)\Gamma\left(\frac{d}{2}-r\right)}
{\Gamma(d+2-s-r)}\enspace k^{m}k^{n}
\biggr],\\
J(k,m_1,m_2)&=&\int\frac{d^3 p}{(2\pi)^3}\frac{1}{\left((p+k)^2+m_{1}^{2}
\right)\left(p^2+m_{2}^{2}\right)}=\frac{A}{8\pi},\\
J^{m}(k,m_1,m_2)&=&\int\frac{d^3
p}{(2\pi)^3}\frac{p_{m}}{\left((p+k)^2+m_{1}^{2}
\right)\left(p^2+m_{2}^{2}\right)}\nonumber\\
&&=\frac{k^{m}}{8\pi}\biggl[\frac{m_{1}-m_{2}}{k^2}-\frac{k^2+m_{1}^{2}
-m_{2}^{2}}{2k^2}\,A\biggr],\\
J^{mn}(k,m_1,m_2)&=&\int\frac{d^3 p}{(2\pi)^3}\frac{p^{m}p^{n}}{\left((p+k)^2+
m_{1}^{2}\right)\left(p^2+m_{2}^{2}\right)}\\
&=&-\frac{\delta^{mn}}{8\pi}\left[\frac{m_1}{2}+\frac{4k^2m_{2}^{2}+
\left(k^2+m_{1}^{2}-m_{2}^{2}\right)^2}{8k^2}\,A\right]\nonumber\\
&&-\frac{k^{m}k^{n}}{8\pi}\left[\frac{m_1}{2k^2}-\frac{4k^2m_{2}^{2}
+3\left(k^2+m_{1}^{2}-m_{2}^{2}\right)^2}{8k^4}\,A\right]\nonumber\\
&&+\frac{1}{8\pi}\frac{(m_{1}-m_2)\left(k^2+m_{1}^{2}-m_{2}^{2}\right)}{4k^2}
\left[\delta^{mn}-3\,\frac{k^{m}k^{n}}{k^2}\right],
\eea
where $A=\frac{2}{k}\arctan\left(\frac{k}{m_1+m_2}\right)$
\bea
j(m)&=&\int\frac{d^3 p}{(2\pi)^3}\frac{1}{p^2+m^2}=-\frac{m}{4\pi},\\
l^{mn}(m)&=&\int\frac{d^3 p}{(2\pi)^3}\frac{p_{m}p_{n}}{p^2+m^2}=
-\frac{\delta^{mn}}{3}m^2j(m).
\end{eqnarray}
These integrals were evaluated using dimensional regularization. 

The diagrams contributing to the 2-point functions of $A_i$ are the
following:

\vskip0.15 truecm
\hspace{2.4cm}
\epsfbox{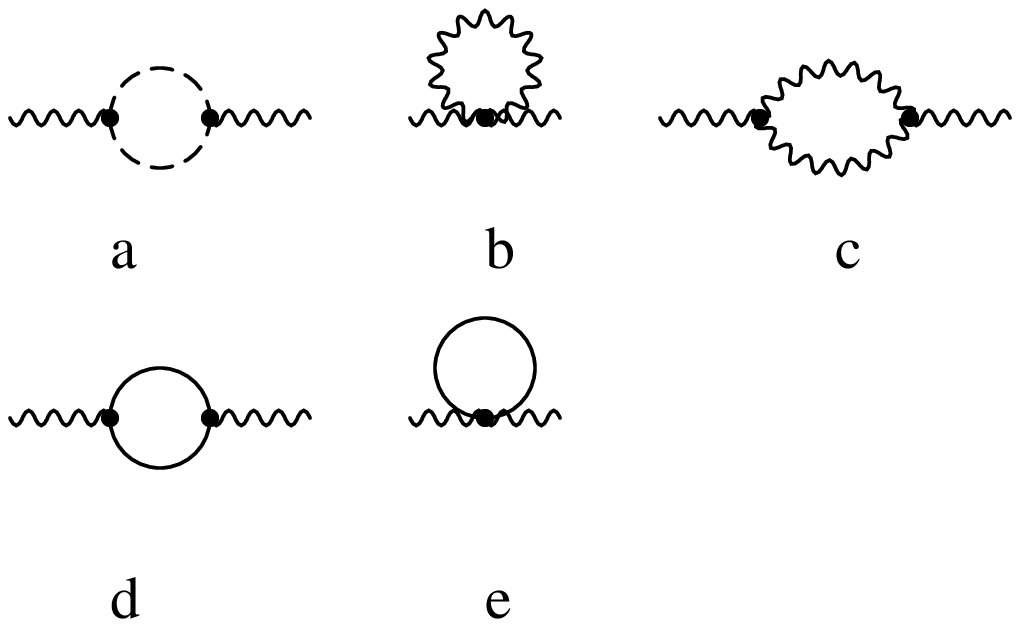}
\vskip0.1truecm
with the analytical contribution:

\begin{eqnarray}
{\Pi^{(a)}}^{mn}_{ab}(k)&=&g^2N\delta_{ab}\biggl[J_{m n}(k,m_G,m_G)
+k_{m}J_{n}(k,m_G,m_G)\biggr],\\
{\Pi^{(b)}}^{mn}_{ab}&=&g^2N\delta_{ab}\delta^{mn}\biggl[\frac{4}{3}j(m_T)+
\frac{2}{3}\xi j(m_L)\biggr],\\
{\Pi^{(c)}}^{mn}_{ab}(k)&=&-\frac{1}{2}g^2N\delta_{ab}
\biggl\{
\delta_{mn}\biggl[-\frac{\left(m^{2}_{T}+k^2\right)^2}{m_{T}^{2}}
\left(J(k,m_T,0)+J(k,0,m_T)\right)\nonumber\\
&&\quad+\frac{2k^2\left(4m^{2}_{T}+k^2\right)}{m^{2}_{T}}J(k,m_T,m_T)
-2\,\left(1+\frac{k^2}{m^{2}_{T}}\right)j(m_{T})\biggr]\nonumber\\
&&\quad+k_{m}k_{n}\biggl[-\frac{\left(m^{2}_{T}+k^2\right)^2}
{4m_{T}^{4}}J(k,m_{T},0)
+\left(\frac{k^4}{4m^{4}_{T}}-\frac{k^2}{m^{2}_{T}}-6\right)J(k,m_T,m_T)
\nonumber\\
&&\quad+\left(-\frac{k^4}{4m^{4}_{T}}+\frac{3}{2}\,\frac{k^2}{m^{2}_{T}}+
\frac{7}{4}
\right)J(k,0,m_T)+\frac{3}{2m^{2}_{T}}j(m_T)+\frac{k^4}{4m^{4}_{T}}I(k,1,1)
\biggl]\nonumber\\
&&\quad+\biggl[-\frac{k_{m}k_{n}}{m^{2}_{T}}j(m_T)+\frac{k^4}{4m^{4}_{T}}
I_{mn}(k,1,1)+\left(\frac{k^4}{4m^{4}}+4\frac{k^2}{m^{2}_{T}}+8\right)
J_{mn}(k,m_T,m_T)\nonumber\\
&&\quad-\frac{2}{m_{T}^{2}}l_{mn}(m_T)-\left(1+\frac{k^2}{m^{2}_{T}}\right)^2
\left(J_{mn}(k,0,m_T)+J_{mn}(k,m_T,0)\right)\biggr]\nonumber\\
&&\quad+\biggl[-\frac{k_{m}k_{n}}{m_{T}^{2}}j(m_T)+\frac{k^4}{m^{4}_{T}}
k_mI_{n}(k,1,1)-
\left(\frac{k^2}{m^{2}_{T}}+1\right)\left(\frac{k^2}{m^{2}_{T}}+3\right)k_m
J_{n}(k,m_T,0)\nonumber\\
&&\quad+\left(1-\frac{k^4}{m^{4}_{T}}\right)k_{m}J_{n}(k,0,m_T)+\left(
\frac{k^4}{m^{4}_{T}}+4\frac{k^2}{m^{2}_{T}}+8\right)k_{m}J_{n}(k,m_T,m_T)
\biggr]
\biggr\}\nonumber\\
&&-g^2N\delta_{ab}\xi\
\biggl\{
\delta_{mn}\biggl[\frac{\left(m_{T}^{2}+k^2\right)^2}{m^{2}_{L}}\left(
J(k,0,m_T)-J(k,m_L,m_T)\right)\nonumber\\
&&\qquad\qquad\qquad\qquad+\frac{\left(m_{T}^{2}+m_{L}^{2}+k^2\right)}
{m^{2}_{L}}j(m_L)
\biggr]\nonumber\\
&&\quad+\frac{k_{m}k_{n}}{4m_{T}^{2}m_{L}^{2}}\biggl[\left(6k^2m_{T}^{2}+
7m_{T}^{4}+
2m_{T}^{2}m_{L}^{2}-\left(m_{L}^{2}+k^2\right)^2\right)J(k,m_L,m_T)\nonumber\\
&&\qquad\qquad\qquad+m_{L}^{2}j(m_T)-7m_{T}^{2}j(m_L)+\left(k^4-6k^2m_{T}^{2}
-7m^{4}_{T}\right)J(k,0,m_T)\nonumber\\
&&\qquad\qquad\qquad-k^4I(k,1,1)+\left(m_{L}^{2}+k^2\right)^2J(k,m_L,0)\biggr]
\nonumber\\
&&\quad+\frac{1}{m_{L}^{2}m_{T}^{2}}\biggl[m_{T}^{2}k_{m}k_{n}j(m_L)
+k^4\left(J_{mn}(k,m_L,0)-I_{mn}(k,1,1)\right)\nonumber\\
&&\qquad\qquad\qquad+\left(k^2+m_{T}^{2}\right)^2\left(J_{mn}(k,0,m_T)
-J_{mn}(k,m_L,m_T)\right)+m_{T}^{2}l_{mn}(m_L)\biggr]\nonumber\\
&&\quad+\biggl[-\frac{\left(k^2+m_{T}^{2}\right)\left(k^2-m_{T}^{2}+
m_{L}^{2}\right)}
{m_{L}^{2}m_{T}^{2}}k_mJ_n(k,m_L,m_T)
-\frac{k^4}{m_{L}^{2}m_{T}^{2}}k_mI_n(k,1,1)\nonumber\\
&&\quad\,\,\,+\frac{k_mk_n}{m_{L}^{2}}j(m_L)+\frac{k^2}{m^{2}_{T}}
\left(\frac{k^2}{m^{2}_{L}}+1\right)k_mJ_n(k,m_L,0)
+\frac{k^4-m_{T}^{4}}{m_{T}^{2}m_{L}^{2}}k_mJ_n(k,0,m_T)\biggr]\biggr\}
\nonumber\\
&&-\frac{1}{2}N\delta_{ab}\xi^2
\biggl\{
\frac{k_mk_n}{4m_{L}^{4}}\biggl[-2m_{L}^{2}j(m_L)+k^4\left(I(k,1,1)+
J(k,m_L,m_L)
\right)\nonumber\\
&&\qquad\qquad\qquad-\left(k^2+m_{L}^{2}\right)^2J(k,m_L,0)-\left(k^2-m_{L}^{2}
\right)^2J(k,0,m_L)\biggr]\nonumber\\
&&\quad+\frac{k^4}{m_{L}^{4}}\biggl[I_{mn}(k,1,1)-J_{mn}(k,m_L,0)-
J_{mn}(k,0,m_L)
+J_{mn}(k,m_L,m_L)\biggr]\nonumber\\
&&\quad+\biggl[-\frac{k^2}{m^{2}_{L}}\left(\frac{k^2}{m^{2}_{L}}+1\right)
k_mJ_n(k,m_L,0)
-\frac{k^2}{m^{2}_{L}}\left(\frac{k^2}{m^{2}_{L}}-1\right)k_mJ_n(k,0,m_L)
\nonumber\\
&&\qquad\qquad\qquad\qquad+\frac{k^4}{m^{4}_{L}}\left(k_mJ_n(k,m_L,m_L)+
k_nI_m(k,1,1)\right)\biggr]\biggr\},\\
{\Pi^{(d)}}^{mn}_{ab}(k)&=&-\frac{1}{2}g^2N\delta_{ab}\biggl[
4J^{mn}(k,m_D,m_D)+4k^m J^{n}(k,m_{D},m_{D})+k^m k^n J(k,m_D,m_D)\biggr],\\
{\Pi^{(e)}}^{mn}_{ab}&=&g^2N\delta_{ab}\delta^{mn}j(m_D).
\end{eqnarray}
Diagrams a), b) and c) were calculated in \cite{kajantiea} using the Landau
gauge ($\xi=0$). The results of these calculations coincide with ours if in the 
above formulas one sets $\xi=0$.
The diagrams contributing to the 2-point function of $A_0$ are given by
diagrams f) and g): 
\vskip0.3truecm
\hspace{4.5cm}
\epsfbox{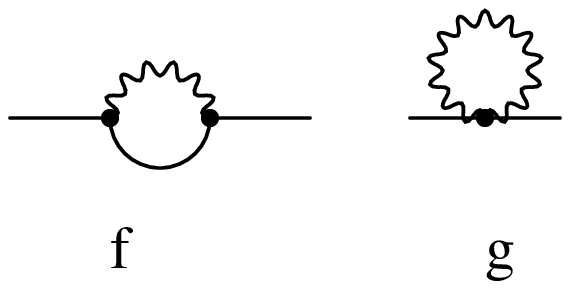}
\vskip0.3truecm
The corresponding analytical contribution is:
\begin{eqnarray}
{\Pi^{(f)}}^{\mu \nu}_{ab}(k)&=&-g^2N\delta_{ab}\delta^{0\mu}\delta^{0\nu}
\biggl[j(m_T)-\frac{k^2+m_{D}^{2}}{m^{2}_{T}}j(m_T)-\frac{\left(m_{D}^{2}+k^2
\right)^2}{m^{2}_{T}}J(k,m_D,0)\\
&&\quad+\frac{1}{m_{T}^{2}}\left(2k^2\left(m_{T}^{2}+m_{D}^{2}
\right)^2+k^4+
\left(m_{T}^{2}-m_{D}^{2}\right)^2\right)J(k,m_D,m_T)-j(m_D)
\biggr]\nonumber\\
&&-g^2N\delta_{ab}\delta^{0\mu}\delta^{0\nu}\frac{\xi}{m_{L}^{2}}
\biggl[
\left(m_{L}^{2}+m_{D}^{2}+k^2\right)j(m_L)+\left(m_{D}^{2}+k^2\right)^2J(k,m_D,0)
\nonumber\\
&&\qquad\qquad-\left(m_{D}^{2}+k^2\right)^2J(k,m_D,m_{L})
\biggr],\nonumber\\
{\Pi^{(g)}}^{\mu\nu}_{ab}&=&g^2N\delta_{ab}\delta^{0\mu}\delta^{0\nu}
\biggl[2j(m_T)+\xi j(m_L)\biggr].
\end{eqnarray}
Finally the single diagram contributing to the ghost 2-point function is
given by diagram h):
\vskip0.3truecm
\hspace{6cm}
\epsfbox{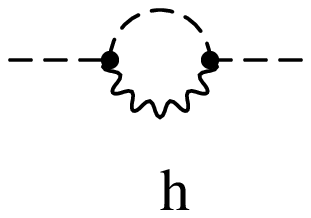}
\vskip0.3truecm
The corresponding analytical contribution is: 
\begin{eqnarray}
{\Sigma}_{ab}(k) &=&-\frac{1}{2}g^2N\delta_{ab}\biggl\{
\biggl[\frac{m_{T}^{2}-\left(k^2+m_{G}^{2}\right)}{4m_{T}^{2}}j(m_T)
-\frac{\left(k^2+m_{G}^{2}\right)^2}{4m_{T}^{2}}J(k,m_G,0)\\\nonumber
&&\qquad\qquad+\left(\frac{\left(k^2+m_{G}^{2}\right)^2}{4m^{2}_{T}}+
\frac{k^2-m_{G}^{2}}{2}+\frac{m_{T}^{2}}{4}\right)J(k,m_G,m_T)
-\frac{1}{4}j(m_G)\biggr]\\\nonumber
&&-\xi\biggl[\left(\frac{\left(k^2+m_{G}^{2}\right)^2}{4m^{2}_{L}}+
\frac{k^2+m_{G}^{2}}{2}+\frac{m_{L}^{2}}{4}\right)J(k,m_G,m_L)
-\frac{1}{4}j(m_G)\\\nonumber
&&\qquad\qquad-\frac{\left(k^2+m^{2}_{G}\right)^2}{4m^{2}_{L}}J(k,m_G,0)
-\frac{3m_{L}^{2}+k^2+m_{G}^{2}}{4m^{2}_{L}}j(m_L)
\biggr]\biggr\}.\\\nonumber
\end{eqnarray}

\end{document}